\documentclass[article,prb,superscriptaddress,amsmath,twocolumn,amssymb]{revtex4-2}

\usepackage{graphicx}
\usepackage{modiagram}
\usepackage{braket}
\usepackage{ulem}   
\usepackage{dsfont}
\usepackage{amsthm,amsmath,amsfonts,amssymb,verbatim,color,tabularx}
\usepackage{bbold}
\usepackage{mathtools}
\usepackage[T1]{fontenc}
\usepackage[colorlinks=true,citecolor=blue,linkcolor=blue,urlcolor=blue]{hyperref}
\usepackage{capt-of}

\usepackage{array, booktabs}
\usepackage[demo,export]{adjustbox}  

\newcommand{\bs}[1]{{\boldsymbol{#1}}}

\newcommand{\pfrms}{\bs{F}^{\text{rms}}_{\text{proj}}}
\newcommand{\pfmax}{max$(\bs{F}^{ij}_{\text{proj}})$}
\begin{document}
\title{Nudged Elastic Membranes for Constructing Reduced Two-Dimensional Potential Energy Surfaces}
\author{Uday Sankar Manoj }
\affiliation{Department of Chemistry, Pennsylvania State University, University Park, Pennsylvania 16802, USA}

\author{Nicole Drew}
\affiliation{Department of Physics, Pennsylvania State University, University Park, Pennsylvania 16802, USA}

\author{Ismaila Dabo}
\affiliation{Department of Materials Science and Engineering, Carnegie Mellon University, 5000 Forbes Avenue, Pittsburgh, PA, 15213, USA}
\author{Lukas Muechler}
\affiliation{Department of Chemistry, Pennsylvania State University, University Park, Pennsylvania 16802, USA}
\affiliation{Department of Physics, Pennsylvania State University, University Park, Pennsylvania 16802, USA}
\email{lfm5572@psu.edu}

\begin{abstract}
Path optimization methods have been widely used and highly successful for the analysis of chemical reactions. Yet, they can fail to capture intrinsically multidimensional features of potential energy surfaces (PES). 
We introduce the nudged elastic membrane method, a framework for constructing two-dimensional reduced potential-energy surfaces in chemically relevant regions of a PES using only energies and forces without requiring more costly Hessian information. 
The method is demonstrated for a three-dimensional prototype model and for the triplet formaldehyde molecular system.
In both cases, the resulting membrane captures one-dimensional reaction-path features as well as genuinely two-dimensional structures such as a yet unreported reported second-order saddle point in the PES of triplet formaldehyde. 
The method further provides direct access to structures that can serve as starting points for subsequent refinement. 
Our results show that the method offers a practical route to exploring multidimensional PES topography beyond the single-path picture.
\end{abstract}
\date\today
\maketitle

\section{Introduction}

Theoretical modeling of chemical reactions is commonly framed as the exploration of a potential energy surface (PES), on which reaction mechanisms are often analyzed using path-based concepts such as transition states (TS), minimum-energy paths (MEP), and intrinsic reaction coordinates (IRCs) \cite{cerjan1981finding,quapp1984analysis,fukui1970formulation}. 
These ideas have been highly successful in describing a wide range of chemical reactions and have motivated a broad range of methods for locating and analyzing reaction pathways, including the nudged elastic band, string, and growing string methods \cite{jonsson1998nudged,henkelman1999dimer,peters2004growing,Schlegel2003PESOverview,DewyerArguellesZimmerman2018ReactionSpace}.

At the same time, 
Notwithstanding its broad utility, a single MEP does not always provide an adequate description of chemical reactivity. Reaction dynamics may deviate from a single path-like picture, and one-dimensional reaction paths can fail to resolve intrinsically multidimensional features of the PES. 
Important examples include roaming dynamics, post-transition-state bifurcations, and higher-order saddle structures, which result in the breakdown of a single-MEP framework both qualitatively and quantitatively \cite{Suits2020Roaming,HareTantillo2017PTSB,VidossichDeVivo2021FirstPrinciplesCatalysis,Ezra2009-mq,Rehbein2011-bk,Hong2014-oy, Pradhan2019-qc, Ess2008-jv,Heidrich1986-pi,Yadav2022-pw}. 

A central difficulty in advancing beyond a one-dimensional path-based description is adequately selecting a reduced coordinate that still captures the relevant reactive motion
For simple systems, or for small regions of the PES, it may be possible to represent the landscape by scanning one or two internal coordinates. As the motion becomes more collective, or as larger regions of the PES are of interest, scan-based approaches fail. A related strategy is then to describe reactivity in terms of collective variables and reduced free-energy surfaces rather than explicit low-dimensional scans~\cite{Yang2019EnhancedSampling,Mehdi2024EnhancedSamplingML,AwasthiNair2019HighDimFES,Piccini2022AIMDEnhancedSampling,VidossichDeVivo2021FirstPrinciplesCatalysis}. These approaches are powerful, but they require a suitable choice of collective variables and typically target free-energy landscapes at finite temperature rather than reduced representations of the underlying PES itself. Moreover, obtaining converged free-energy information at finite temperature generally requires extensive sampling, which can remain costly even for relatively small systems. Thus, for applications in which one seeks a direct, geometrically resolved representation of a chemically relevant region of the PES from less costly electronic-structure energies and forces, a more explicit surface-construction framework remains desirable.

Previous work has approached this problem by interpolating between multiple IRCs~\cite{chuang2020construction}. 
However, a general framework for constructing and analyzing multidimensional reactive regions of the PES in absence of IRCs or larger regions of the PES is still lacking. 

In this work, we introduce the nudged elastic membrane (NEM) method, a formalism extending beyond the nudged elastic band approach, to construct a two-dimensional manifold in coordinate space that captures a chemically relevant region of the PES. The resulting membrane is designed to resolve multiple reaction pathways and higher-order features such as second-order saddle points (SOSP) within a single reduced representation. In the following, we present the conceptual and numerical framework of NEM, describe its implementation and optimization, and illustrate its performance on two representative examples, namely a three-dimensional prototype model and the triplet formaldehyde system.
 
\section{Conceptual Definitions}
\subsection{Nudged Elastic Membrane}
\begin{figure*}[t]
    \centering
    \includegraphics[width=0.85\textwidth]{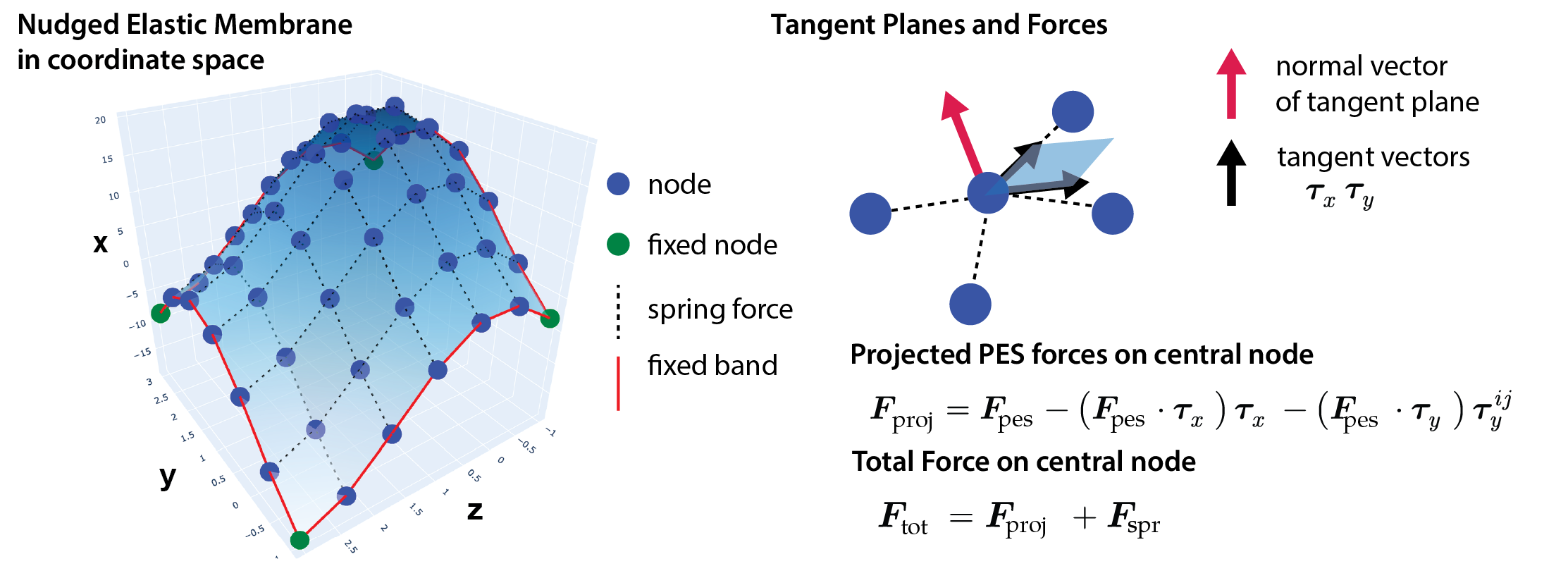}
    \caption{(a) A representative illustration of an NEM in three dimensions. The $x$-, $y$-, and $z$-axis represent degrees of freedom used to describe the system. (b)  Tangent vectors $\bs{t}_x$, $\bs{t}_y$(black arrows) and  orthorgonal projected force vector $\bs{F}_{\text{proj}}$ (red arrow) on a central node.}
    \label{fig:grid}
\end{figure*}
We begin by outlining the general concepts underlying the NEM method. The NEM can be conceptualized as an elastic manifold embedded in a coordinate space that conforms to the underlying PES, revealing its shape and structural features. 
It is represented by a grid of $N_x \times N_y$ beads in the space of nuclear configurations [Fig.~\ref{fig:grid}]. 
Each bead $\bs{R}_{ij} \  (i \leq N_x$,$j \leq N_y)$ corresponds to a specific molecular geometry, and is connected to four nearest neighbors via fictitious springs. In addition to spring forces, each bead experiences forces derived from the PES $\bs{F}^{ij}_{\text{PES}} = -\nabla V(\bs{R}_{ij})$, where $V(\bs{R}_{ij})$ is the total molecular energy with a fixed nuclear configuration $\bs{R}_{ij}$.

The central objective is to minimize the forces on each bead
\begin{equation}
\label{eq:total_force}
    \bs{F}^{ij}_{\text{tot}}=  \bs{F}^{ij}_{\text{proj}} + \bs{F}^{ij}_{\text{spr}},
\end{equation}
which consists of spring forces $\bs{F}^{ij}_{\text{spr}}$ acting in the tangent plane of each bead and projected PES forces $\bs{F}^{ij}_{\text{proj}}$, in which tangent plane components have been removed. \\

The the spring force on bead $\bs{R}_{ij}$ is given by
\begin{equation}
\begin{split}
        \bs{F}^{ij}_{\text{spr}}/k  &= \left( \lVert \bs{R}_{i+1,j} - \bs{R}_{ij}\rVert - \lVert\bs{R}_{i-1,j} - \bs{R}_{ij}\rVert \right) \bs{t}^{ij}_x  \\ &+ \left( \lVert \bs{R}_{i,j+1} - \bs{R}_{ij}\rVert - \lVert\bs{R}_{i,j-1} - \bs{R}_{ij}\rVert \right) \bs{t}^{ij}_y \ ,
\end{split}
\end{equation}
where $k$ is a spring constant, assumed to be independent of the beads for simplicity.
The spring forces depend on the tangent vectors $\bs{t}^{ij}_x$ and $\bs{t}^{ij}_y$ in analogy to the tangent vectors in nudged elastic band theory (NEB)~\cite{henkelman2000improved,jonsson1998nudged}, as illustrated in Fig.~\ref{fig:grid}. They will be defined in detail the following section.

The projected PES forces are given by,

\begin{equation}
\begin{split}
    \boldsymbol{F}^{ij}_{\text{proj}} = \boldsymbol{F}^{ij}_{\text{PES}} - \left(\boldsymbol{F}^{ij}_{\text{PES}} \cdot \boldsymbol{\tau}^{ij}_x \right) \boldsymbol{\tau}^{ij}_x\\ \quad- \left(\boldsymbol{F}^{ij}_{\text{PES}} \cdot \boldsymbol{\tau}^{ij}_y \right) \boldsymbol{\tau}^{ij}_y, 
    \end{split}
\end{equation}
where $\boldsymbol{\tau}^{ij}_x $ and $\boldsymbol{\tau}^{ij}_y$ are an orthonormal basis in the tangent plane at node $\bs{R}_{ij}$.
The spring forces lie entirely along the local tangent plane, and only depend on absolute distance between beads~\cite{henkelman2000improved,jonsson1998nudged}. This ensures that upon minimization of $\bs{F}^{ij}_{\text{tot}}$, $\bs{F}^{ij}_{\text{spr}}$ are minimized, resulting in an equispaced membrane with negligible orthogonal force.

\section{Numerical Implementation}
\begin{figure*}[tb]
    \centering
    \includegraphics[width=0.8\linewidth]{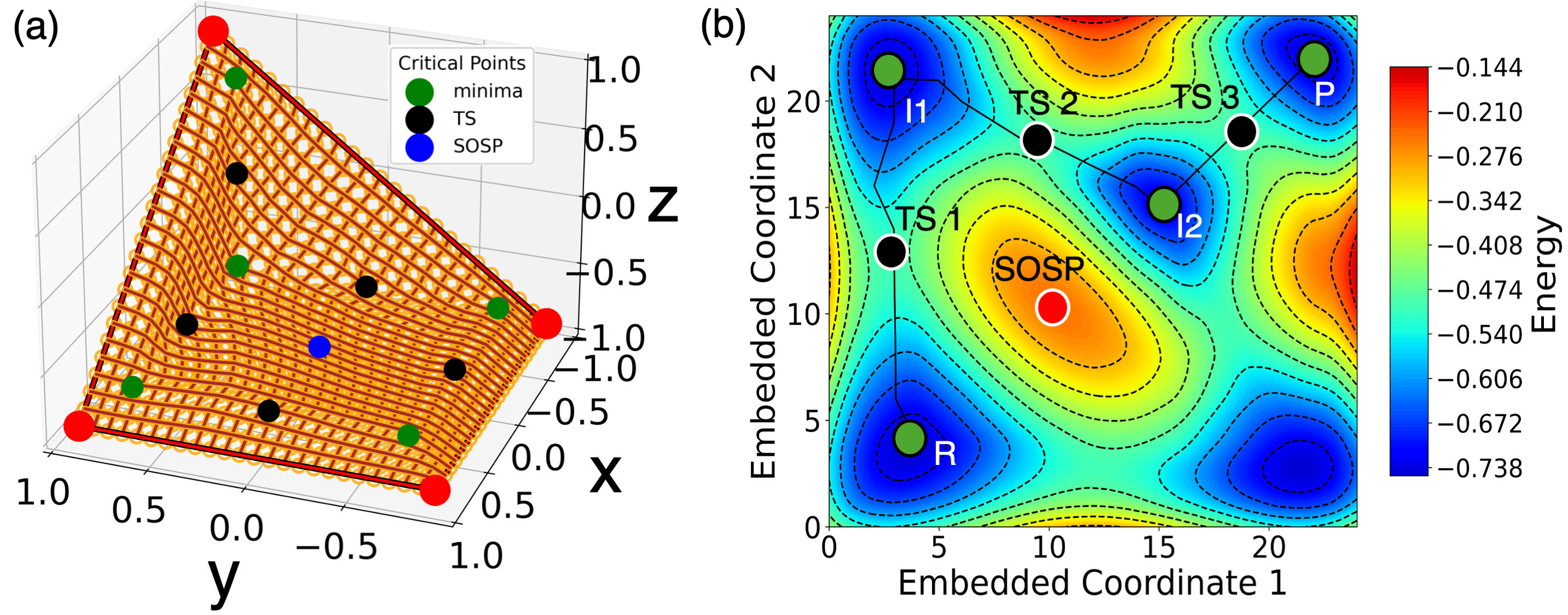}
    \caption{(a) Optimized $25 \times 25$ NEM grid. Green, black, and red points indicate minima, TS, and SOSP, respectively. The grid is concentrated around these critical points, encompassing the low-energy regions of the potential energy surface (PES). (b) The projection of the MEP (black line) on the optimized $25\times25$ grid energy surface. Here, the projection of minima (green dot), TS (black dot), and SOSP (red dot) lie along the valleys on the energy surface. }
    \label{fig:Upscal_opt_grid}
\end{figure*}
\subsection{Tangent Vectors}
The tangent vectors ($\bs{t}^{ij}_x$ and $\bs{t}^{ij}_y$) define the direction of the spring connections in the $3N$ dimensional cartesian coordinate space, for an N atoms system.
$\bs{F}^{ij}_{\text{spr}}$ acts along these tangents and ensures equispacing of beads.
The most straightforward implementation for a tangent vector is a two point tangent approximation used in earlier formulations of NEB,
\begin{equation}
\begin{split}
    & \bs{t}^{ij}_x = \frac{R_{i+1,j} - R_{i-1,j}}{\lVert R_{i+1,j} - R_{i-1,j}\rVert}, \\
    & \bs{t}^{ij}_y = \frac{R_{i,j+1} - R_{i,j-1}}{\lVert R_{i,j+1} - R_{i,j-1}\rVert}
\end{split}
\end{equation}
However, it is also well known from NEB that this formulation can lead to kinks, since the spring forces have no component orthogonal to the NEB path.
Therefore, we here use a formalism analogous to the one defined in the work by Henkelman \textit{et al.} \cite{henkelman2000improved}, which uses an energy weighted methodology.
Here, the tangent at each bead is defined based on the relative energy with respect to neighboring beads.
If the bead $\bs{R}_{ij}$ is a local energy extremum with respect to its nearest neighbors, then an energy-weighted linear combination of the forward and backward tangents is chosen as the tangent vector.
Otherwise, if a bead is not a local energy extremum, then the tangent pointing towards the higher energy bead is taken.
Concretely, we define forward and backward tangents at bead $\bs{R}_{ij}$.
\begin{align}
    \bs{t}^{ij}_{+x} &= \bs{R}_{i+1,j} - \bs{R}_{ij}  \\
    \bs{t}^{ij}_{-x} &= \bs{R}_{ij} - \bs{R}_{i-1,j}
\end{align}
The forward and backward tangents along $y$-direction, $\bs{t}_{+y}$ and $\bs{t}_{-y}$, are defined analogously.

If the bead $\bs{R}_{ij}$ is a not an extremum along the $x$-direction (or $y$-direction), the tangent is defined as:
\begin{equation}
    \bs{t}^{ij}_x  =
    \begin{cases}
        \bs{t}^{ij}_{+x} & \text{if $V_{i+1,j} >V_{i,j} >V_{i-1,j}$ } \\
        \bs{t}^{ij}_{-x} & \text{if $V_{i+1,j} <V_{i,j} <V_{i-1,j}$ } 
    \end{cases}
\end{equation}

If the bead $\bs{R}_{ij}$ is a local extremum, $\bs{t}^{ij}_x$ is defined as:
\begin{equation}
    \bs{t}^{ij}_x  =
    \begin{cases}
        \bs{t}^{ij}_{+x}\Delta V^x_{\text{max}} + \bs{t}^{ij}_{-x}\Delta V^x_{\text{min}} & \text{if $V_{i+1,j}>V_{i-1,j}$} \\
        \bs{t}^{ij}_{+x}\Delta V^x_{\text{min}} + \bs{t}^{ij}_{-x}\Delta V^x_{\text{max}} & \text{if $V_{i+1,j}<V_{i-1,j}$}
    \end{cases}
\end{equation}
Where $\Delta V^x_{\text{max}}$ and $\Delta V^x_{\text{min}}$ are defined as
\begin{align}
    \Delta V^x_{\text{max}}&= \text{max}(|V_{i,j} - V_{i+1,j}|,|V_{i,j} - V_{i-1,j}|) \\
    \Delta V^x_{\text{min}}&= \text{min}(|V_{i,j} - V_{i+1,j}|,|V_{i,j} - V_{i-1,j}|)
\end{align}

Finally, $\bs{t}_x$ ($\bs{t}_y$) is normalized to obtain a unit vector that defines the local $x$- or $y$-direction. The basis vectors spanning the local tangent plane, $\boldsymbol{\tau}^{ij}_x$ and $\boldsymbol{\tau}^{ij}_y$, are then obtained through Gram-Schmidt orthogonalization of $\bs{t}^{ij}_y$ with respect to $\bs{t}^{ij}_x$~\cite{leon2013gram}.

\subsection{Grid Initialization and Optimization}

To initialize the NEM, a grid is initially constructed encompassing a region of interest. 
In the most simple approach four anchoring corner points are chosen, and linear interpolation along one-dimensional cuts is used to define the grid.
Alternatively, if IRCs/MEPs of interest are known, an initial grid can be constructed by interpolating between IRCs/MEPs\cite{chuang2020construction}.

The anchor points and the edges of the grid are kept fixed, and the inner beads are displaced along $\bs{F}^{ij}_{\text{tot}}$ in each iteration. 
In this work, convergence is tracked using the root-mean-square of $\bs{F}^{ij}_{\text{proj}}$ of the inner, non-fixed beads,
\begin{equation}
    \bs{F}^{\text{rms}}_{\text{proj}} = \sqrt{\frac{\sum_{i,j}\lVert \bs{F}^{ij}_{\text{proj}}\rVert^2}{(N_x-2)(N_y-2)}}
\end{equation}
where the summation is over $i={2,3,\cdots N_x-1}$, and $i={2,3,\cdots N_y-1}$.

In this work, an in-house version of the Fast Inertial Relaxation Engine (FIRE) 2.0 in conjugation with a semi-implicit Euler integrator~\cite{bitzek2006structural,guenole2020assessment} was used for the optimization procedure.
FIRE is a momentum-type optimizer with variable time step ($dt$) and velocity-force mixing, controlled through a paramter $\alpha$. Both are dynamically changed by the algorithm itself~\cite{ribaldone2022fast}.
This algorithm can accelerate and decelerate through the dynamic control of $dt$, and has been employed for the optimization of NEB~\cite{herbol2017computational,sheppard2008optimization}.
The FIRE algorithm has several other tunable and fixed parameters. 
For our discussion, the most important parameters are $dt_{\text{in}}$, $\alpha_0$, $f_{\text{inc}}$, $f_{\text{dec}}$, $dt_{\text{max}}$, $N_{\text{del}}$ and $f_{\alpha}$. 
A detailed rationale behind these parameters and their updates can be found in earlier works on FIRE~\cite{ribaldone2022fast,bitzek2006structural,SHUANG2019135} and for each simulation the choices of parameters will be explicitly given.

In our implementation of the FIRE algorithm, we define one extra parameter, $dR_{\text{max}}$, which dictates the maximum allowed step-size along the local tangent plane.
The optimization is considered converged when $\bs{F}^\text{rms}_\text{proj}$ is below a chosen threshold.

In our approach, we found a uniform choice of $k$ to be the most robust.  The value  of $k$ is chosen based on the magnitude of $\bs{F}^{ij}_{\text{proj}}$, i.e. 
\begin{equation}
    k = \frac{\text{max}(\lVert F^{ij}_\text{proj}\rVert)}{2\times dR_{\text{max}}}
    \label{eq:k}
\end{equation}

This ensures the magnitude of $\bs{F}^{ij}_{\text{spr}}$ is close to that of \pfmax, and that the optimization is driven by both $\bs{F}^{ij}_{\text{proj}}$ and $\bs{F}^{ij}_{\text{spr}}$.
For accelerated convergence with respect to the number of beads, we employ an upscaling approach, motivated from earlier improvements on the NEB method~\cite{kolsbjerg2016automated}. 
After the optimization of an intial small grid, a grid upscaling protocol is introduced to increase the resolution of the NEM.
The grid upscaling introduces an extra bead between each spring connection along $x$-, $y$-direction, as well as along the diagonal, effectively doubling the resolution of the inner grid. 
This facilitates a computationally less intensive pre-relaxation of the membrane.

\begin{table*}[t]
    \centering
    \caption{Comparison of exact and NEM-predicted geometries at stationary points for the 3D prototype model  (Eq.~\ref{eq:3dmodel}). $\Delta_{\mathrm{RMSD}}$ denotes the root-mean-square deviation of the geometries, and $\Delta E$ the absolute energy difference between reference and predicted structures [Fig.~\ref{fig:Upscal_opt_grid}(b)].}
    \begin{tabular}{c c c c c}
        \toprule
        Geometry & Exact  & Prediction & $\Delta_{\mathrm{RMSD}}$ & $\Delta E$ \\
        \hline
        R &($-0.707$,  $+0.707$, $-0.707)$ & ($-0.704$, $+0.699$, $-0.734$) & $0.016$ & $0.0016$\\
        I1 &($-0.707$, $-0.707$, $-0.707$) & ($-0.717$, $-0.713$, $-0.729$) & $0.018$ & $0.0016$\\
        I2 &($+0.707$, $-0.707,$ $-0.707$) & ($+0.623$, $-0.643$, $-0.711$) & $0.061$ & $0.0198$\\
        P &($+0.707$, $-0.707$, $+0.707$) & ($+0.713$, $-0.701$, $+0.698$) & $0.007$ & $0.0164$ \\
        TS1 &($-0.707$,  $+0.000$, $-0.707)$ & ($-0.695$, $+0.011$, $-0.708$) & $0.009$ & $0.0001$\\
        TS2 &($+0.000$, $-0.707$, $-0.707$) & ($-0.004$, $-0.689$, $-0.656$) & $0.032$ & $0.0056$\\
        TS3 &($+0.707$, $-0.707,$ $+0.000$) & ($+0.684$, $-0.684$, $-0.005$) & $0.019$ & $0.0020$\\
        SOSP & ($+0.000$, $+0.000$, $-0.707$) &($-0.027$, $+0.026$, $-0.703$) & $0.022$& $0.0014$\\
        \hline 
        Avg. & & & $0.023$ & $0.0057$\\
        \hline
    \end{tabular}
    \label{tab:DobWell_critic_points}
\end{table*}

\subsection{Energy Surface and Projection}
To obtain an energy surface, we plot the total energy as a function of bead indices. 
This way of plotting is justified since the spring forces keep the beads equidistant.
Cubic splines can be used to interpolate between grid points to obtain a smooth continuous surface.
This energy surface is a reduced dimensional relaxed energy surface which can be employed for further investigation of a chemical system\cite{liu_molecular_2019}.

Projection of a point, $\mathbf{p}$, onto the NEM grid is performed by determining the closest approach between the grid and $\mathbf{p}$. The closest grid point to $\mathbf{p}$ is first located, after which the point of closest approach on the grid is refined via linear interpolation using the local plane vectors.

To project the MEP onto the optimized NEM, the continuous MEP is discretized into a set of points, each of which is subsequently projected onto the grid following the above procedure, yielding the projected MEP.

\section{Results}
\subsection{3D-Model}

As a proof of concept, we first investigate an analytical PES depending on three nuclear degrees of freedom $x,y,z$ for which all the stationary points are known.
The potential function is given by: 
\begin{equation}
    V(x,y,z) = x^4 - x^2 + y^4 - y^2 + z^4 - z^2
    \label{eq:3dmodel}
\end{equation}
This potential possesses eight minima, twelve TSs, six SOSPs, and a local maximum. The minima are located at 
\begin{equation}
   \frac{1}{\sqrt{2}} \left(\pm 1, \pm1, \pm1\right).  
\end{equation}
The local maximum is located at the origin, while the SOSP form the faces of a cube. 
The SOSP of interest in this study is located at $(\frac{1}{\sqrt{2}},0,0)$.
A TS connects each pair of adjacent minima, and TSs are interconnected via SOSPs.
This configuration gives rise to multiple MEPs interconnecting the minima, with the SOSP situated between these pathways as shown in Tab.~\ref{tab:DobWell_critic_points}.

We initialize a $7\times7$ grid through linear interpolation between four anchor points located at:
\begin{align*}
\mathbf{r}^{(1)} &= (-r_0,-r_0,-r_0), &
\mathbf{r}^{(2)} &= (-r_0,\phantom{-}r_0,-r_0),\\
\mathbf{r}^{(3)} &= (\phantom{-}r_0,\phantom{-}r_0,-r_0), &
\mathbf{r}^{(4)} &= (\phantom{-}r_0,-r_0,\phantom{-}r_0).
\end{align*}
Here $r_0=\frac{1}{\sqrt{2}}+0.2$. The anchor points were chosen near four different minima, and the goal of the calculation is to capture the connectivity between them, including the SOSP. 

This initial grid is upscaled and optimized multiple times to accelerate convergence with respect to the grid size. 
All optimizations use a convergence criterium of  $\bs{F}^{\text{rms}}_{\text{proj}}=0.05$, with the same FIRE parameters: $f_{\text{inc}}=1.2$, $f_{\text{dec}}=0.5$, $f_{\alpha}=0.9$ and $\alpha_0=1.0$, $N_{\text{del}} =4$, $dt_{\text{ min}}=1\times10^{-4}$. 
Only the initial time step, $dt_{\text{in}}$, and maximum time step, $dt_{\text{ max}}$, are adjusted for each grid size.

The initial $7\times7$ grid  is characterized by $\bs{F}^{\text{rms}}_{\text{proj}}=0.387$ and \pfmax$=0.886$.
The grid parameters $k$ and $dR_{\text{ max}}$ were set to $29.3$ and $0.015$, and the spring constants were kept fixed during the optimization.
The time step parameters used are: $dt_{\text{in}}=0.1$, $dt_{\text{max}}=0.5$.
The convergence criteria $\bs{F}^{\text{rms}}_{\text{proj}}=0.05$, was achieved after 47 steps with  \pfmax$= 0.116$ [Fig.~\ref{figS:3D_dbw_Grid}]. 

After this preliminary optimization, we upscaled to a $13\times13$ grid.
This grid is similarly optimized, with $dt_{\text{in}}=0.05$, $dt_{\text{ max}}=0.1$, and $dR_{\text{max}}=0.007$, which converged after 25 steps,with \pfmax$=0.247$.
The energy surface obtained from these two optimization already clearly feature the expected minima, TSs and the SOSP.
However, to obtain more accurate geometries and energies, the grid is upscaled again and optimized with the same convergence criteria, while lowering $dt_{\text{in}}=0.01$, $dt_{\text{max}}=0.1$, and $dR_{\text{max}}=0.004$ for increased precision. This optimization convergence in 37 steps resulting in a $25\times25$ grid with $\pfrms\approx0.049$, and \pfmax$=0.3626$.

The optimized grid lies along the critical points on the PES and the beads are equidistant [Fig.~\ref{fig:Upscal_opt_grid}(a)].
The corresponding energy surface obtained from the NEM reproduces all expected features of the PES.
Geometries for these points are obtained by selecting the bead closest to each structure on the energy surface and are compared to the exact values in Tab.~\ref{tab:DobWell_critic_points}.
The root mean square difference between the exact and approximate structures is below 0.04 in all cases while staying below 0.02 for most structures. This shows that stationary points identified in the energy surface [Fig.~\ref{fig:Upscal_opt_grid}(b)] are qualitatively and quantitatively correct.

An additional benchmark for NEM grid is to project the true MEP onto the optimized NEM grid.
If the points of interest lie within the NEM grid, the projected MEP should follow the low-energy valleys, traverse the TS region, and connect the relevant minima on the energy surface of the NEM grid.
This is showcased by Fig.~\ref{fig:Upscal_opt_grid}(b), where the MEP connecting the minima (green dots) follows the TSs (black dots) predicted by the energy surface. 

\subsection{Triplet Formaldehyde}
To test the performance of the NEM method for more complex and realistic system, we turn to triplet formaldehyde (H$_2$CO$^3$).
This PES features several energetic minima connected by TSs, which are well established and therefore a meaningful test for our method.
We used a PES (PIP-NN) fitted at CCSD(T)-F12a/AVTZ level of theory, accessed through the \texttt{chempotpy} python library~\cite{shu2023chempotpy,liu_molecular_2019}.
\begin{figure}[htb]
    \centering
    \includegraphics[width=0.9\linewidth]{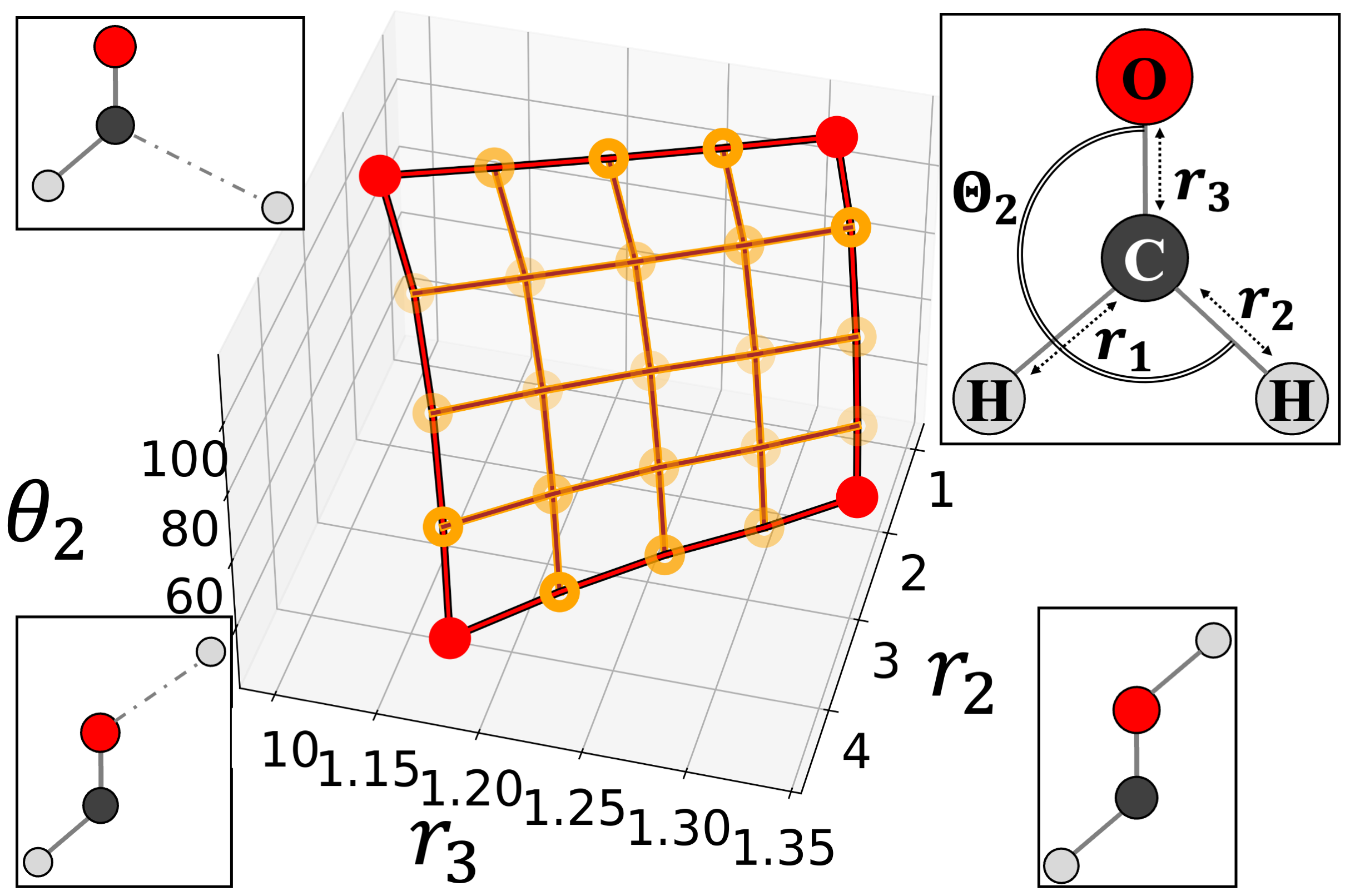}
    \caption{Initial $5\times5$ grid for the formaldehyde system in the $\left(r_2,r_3,\theta_2\right )$ subspace. Red points mark the fixed anchor points connected by fixed red springs; orange springs are freely moving. Schematic molecular geometries at each anchor point are shown in the adjacent black boxes.}
    \label{fig:init_grid}
\end{figure}
\subsubsection{Grid and optimization}
The NEM grid is constructed in cartesian coordinates, with distances in angstroms (\AA) and all forces are evaluated numerically in eV/\AA.
The anchor points for the initial grid are chosen such that it encompasses three known local minima that are important for the formaldehyde decomposition reaction on the lowest triplet surface. 
The first two anchor points correspond to the triplet local minima of formaldehyde (H$_2$CO) and hydroxymethylene (HCOH), which has a hydrogen peroxide-like geometry.
The next two anchor points correspond to two non-equilibrium dissociated structures, one for O-H bond dissociation (from HCOH) and one for C-H bond dissociation (from H$_2$CO).
Linear interpolation is used to construct a $5\times5$ grid from the anchor points, which is illustrated in Fig.~\ref{fig:init_grid} along the $\left(r_2,r_3,\theta_2\right )$ subspace  with a schematic structure of the anchor points. 

Except for $dt_{\text{in}}$ and $dt_{\text{ max}}$, all other FIRE parameters are identical for all optimizations:  $f_{\text{inc}}=1.1$, $f_{\text{dec}} = 0.5$, $\alpha_0=1.0$, $f_{\alpha}=0.9$, and $N_{\text{del}}=5$.
The initial $5\times5$ grid is characterized by $\pfrms = 5.26~\mathrm{eV/\text{\AA}}$ and \pfmax$= 8.01~\mathrm{eV/\text{\AA}}$. 
We set $dR_{\text{max}} = 0.014\text{\AA}$, $k = 291.23~\mathrm{eV/\text{\AA}^2}$, $dt_{\text{in}}=0.1$, and $dt_{\text{max}}=0.4$,

The initial $5\times5$ grid is preconditioned for $50$ iterations lowering $\pfrms$ to $1.51~\mathrm{eV/\text{\AA}}$, and \pfmax$=2.84~\mathrm{eV/\text{\AA}}$.
This grid is unscaled to a $9\times9$ grid, which is further preconditioned for $50$ iterations with $dt_{\text{in}}=0.05$, $dt_{\text{max}}=0.1$, and $dR_{\text{max}}=0.007\text{\AA}$.
This results in a grid with $\pfrms=0.95~\mathrm{eV/\text{\AA}}$ and \pfmax$=2.31~\mathrm{eV/\text{\AA}}$).
The initial, upscaled, and optimized grids are visualized along a few internal coordinates in Fig.~\ref{figS:H2CO_upscale}.
We restricted the number of iterations to $50$, since we found that the $5\times5$ or $9\times9$ grid lacks the resolution to properly sample the large coordinate space, making further optimization cost-ineffective.

\begin{figure}[ht]
    \includegraphics[width=0.95\linewidth]{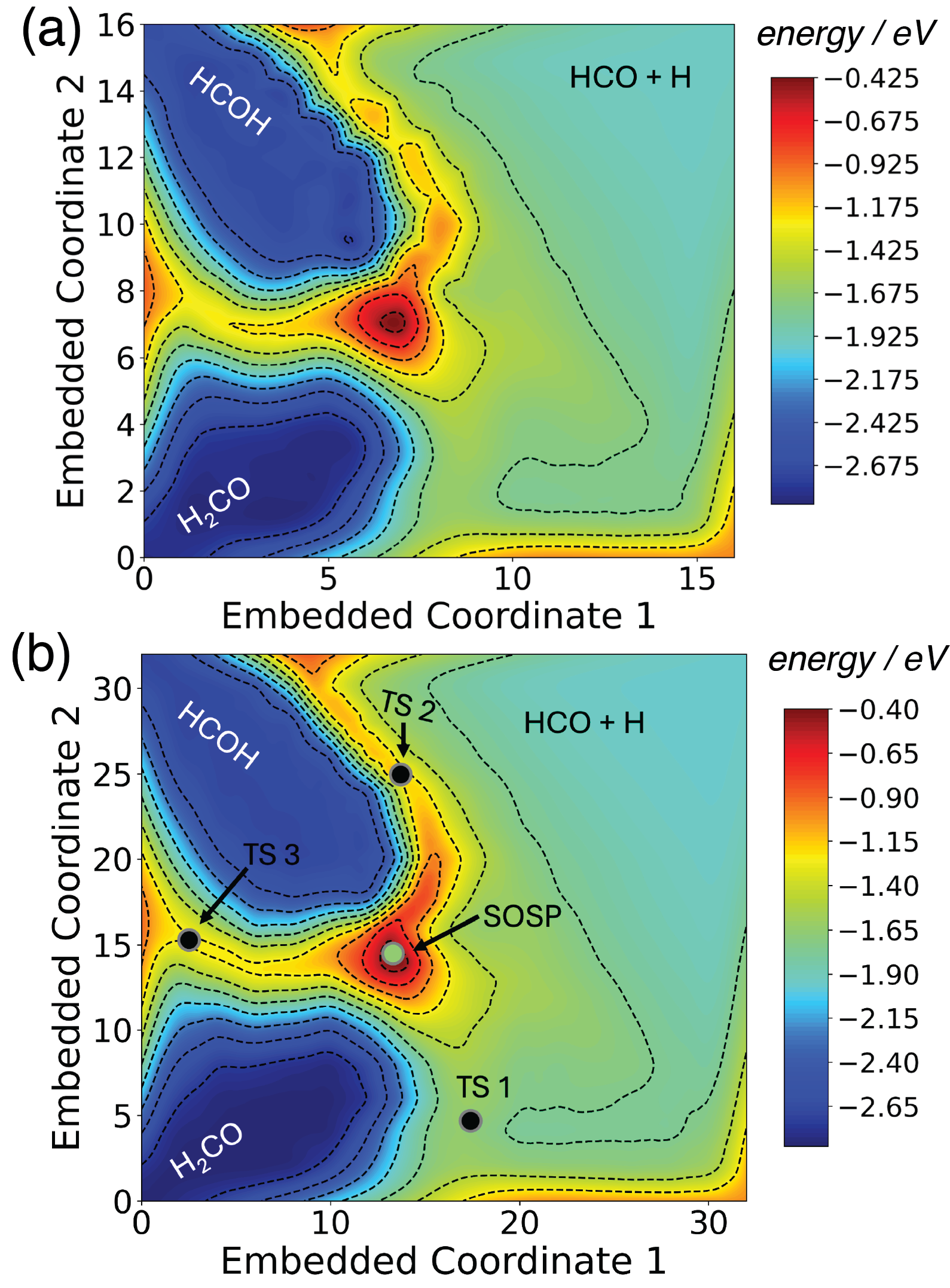}
    \caption{
    (a) Contour plot of the energy surface obtained from the optimized $17\times17$ NEM grid for the formaldehyde system.
    (b) Contour plot of the upscaled energy surface obtained from the optimized $17\times17$ NEM grid for the formaldehyde system. The upscaling process increases resolution over the grid which helps in identifying TS and other PES features. The TSs are marked by black dots and the SOSP is indicated by the green dot.}
    \label{fig:opt_E_form}
\end{figure}

As a last step, the grid is upscaled once more to obtain a $17\times17$ grid [Fig.~\ref{figS:H2CO_upscale_2}], and is optimized with the convergence criterion $\pfrms=0.5~\mathrm{eV/\text{\AA}}$.
The parameters used are $dt_{\text{in}}=0.02$, and $dt_{\text{max}}=0.2$ and $dR_{\text{max}}=0.003\text{\AA}$, and convergence was reached after $189$ steps.
The optimized NEM grid has and a maximum \pfmax$=1.32~\mathrm{eV/\text{\AA}}$ and the energy surface corresponding to this grid is shown in Fig.~\ref{fig:opt_E_form}(a). 

The energy surface  allows identification of the minima corresponding to formaldehyde (bottom left), HCOH (top left), and the broad, shallow valley on the right corresponding to the HCO + H minimum. 
Moreover, we can identify the TS regions connecting these minima. 
To better identify TS regions we do a final upscaling of the optimized grid to obtain a $34\times34$ grid [Fig.\ref{fig:opt_E_form}(b)] which is not optimized further. 
The geometry of approximate TSs retrieved from the NEM grid is compared with the true geometries reported in literature in Tab.~\ref{tab:H2CO_TSs}.
The predicted structures agree well with the reported transition state structures reported in previous studies with a mean RMSD of $0.15 \text{\AA}$ and mean absolute energy error of $\Delta E = 1.485 \frac{\text{kcal}}{\text{mol}}$~\cite{li2017kinetics,hickson2016quantum}.

Additionally, we observe a structure resembling a SOSP on the PES that has not been reported previously. 
To verify this structure, we employed a gradient minimization based on numerical derivatives to verify the existence of a SOSP for this fitted PES, which confirmed the existence of a SOSP with two negative hessian eigenvalues of  $-36.0932096 \frac{\text{eV}}{\text{\AA}^2 \cdot \text{amu}}$ and $-10.7757741 \frac{\text{eV}}{\text{\AA}^2 \cdot \text{amu}}$.
The obtained geometry, RMSD deviation in geometry and energy difference are summarized in Tab.~\ref{tab:H2CO_TSs}.
This highlights the usefulness of the NEM approach, as SOSP are not diagnosable using conventional, one-dimensional methods.

\begin{table*}[tb]
    \caption{Geometry and energy comparisons of the predicted TSs, obtained from the optimized $17\times17$ grid, and the actual TSs from literature. TS1 geometry was reconstructed from the work by Yamaguchi \textit{et al.}\protect\cite{yamaguchi1998unimolecular}. TS2 and TS3 geometries were obtained from the work by Hickson \textit{et al.}\protect\cite{hickson2016quantum}}

    \begin{tabular}{c c c c c }
        \toprule
        Geometry & Predicted Geometry & Actual Geometry & RMSD/\AA & $\Delta E$/kcal mol$^{-1}$ \\
        \midrule
        TS1 &
        \includegraphics[width=0.25\linewidth,valign = c]{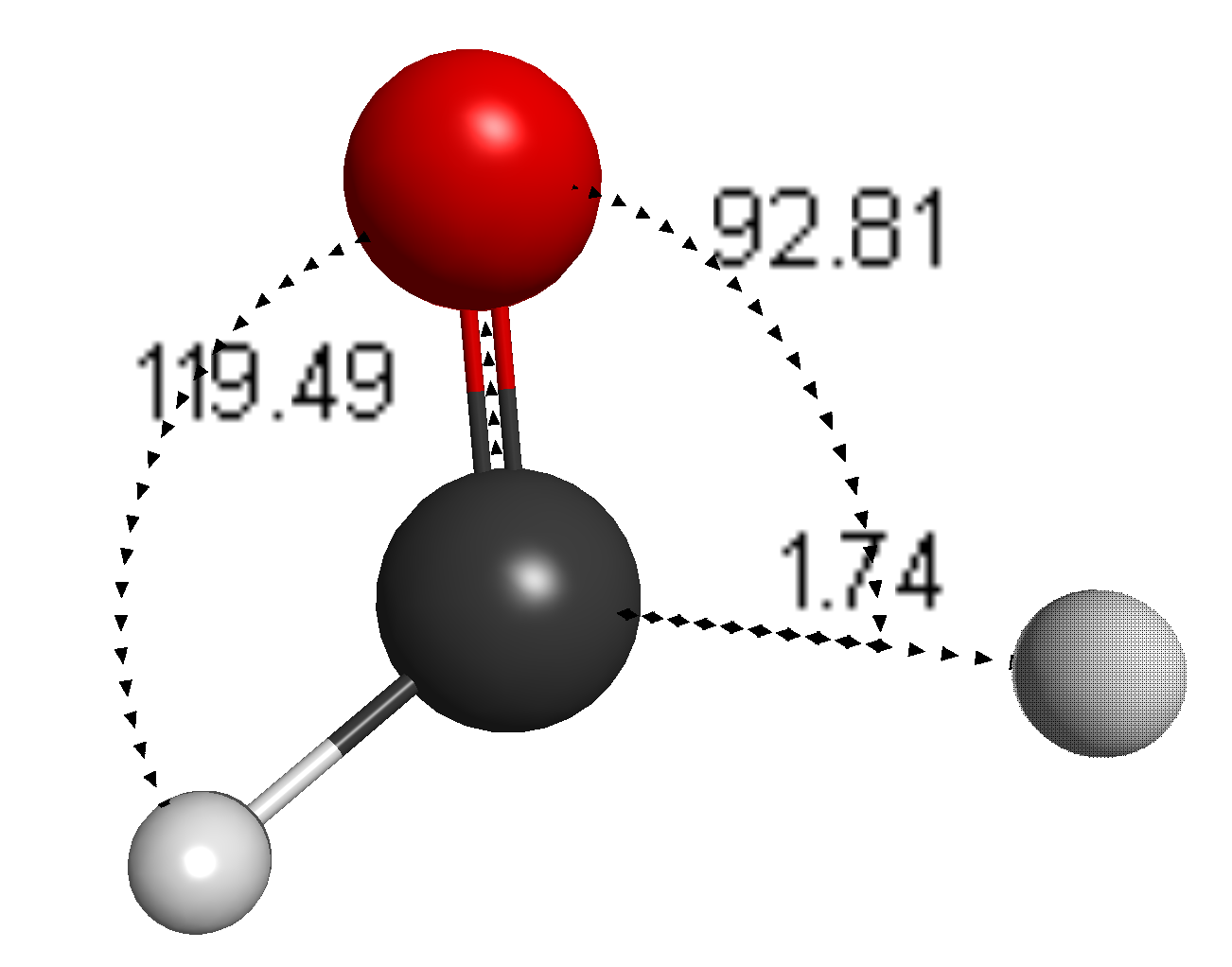} &
        \includegraphics[width=0.2\linewidth,valign = c]{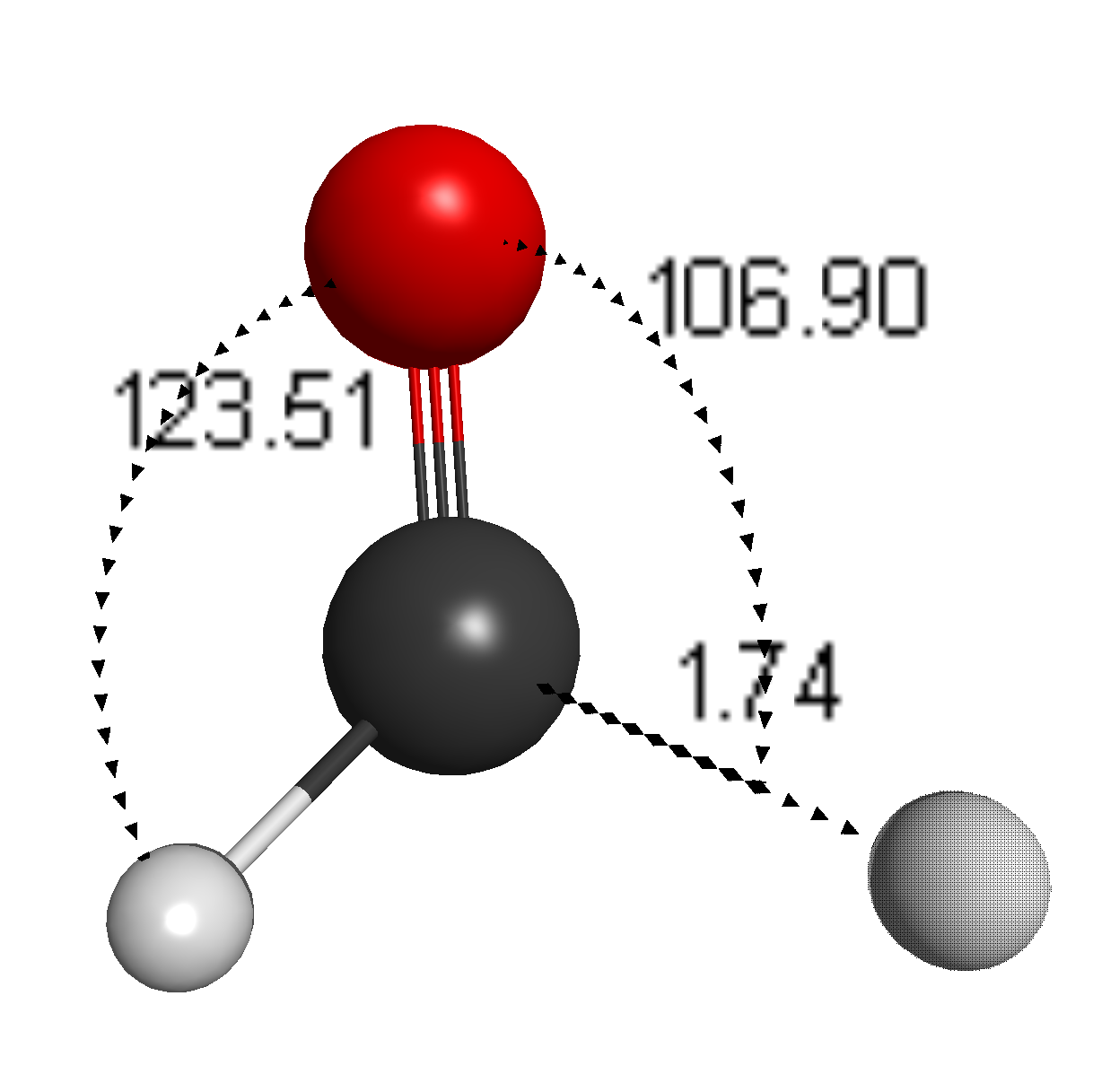} &
        $0.217$ & $2.80$ \\
        TS2 &
        \includegraphics[width=0.25\linewidth,valign = c]{C-H_TS.png} &
        \includegraphics[width=0.25\linewidth,valign = c]{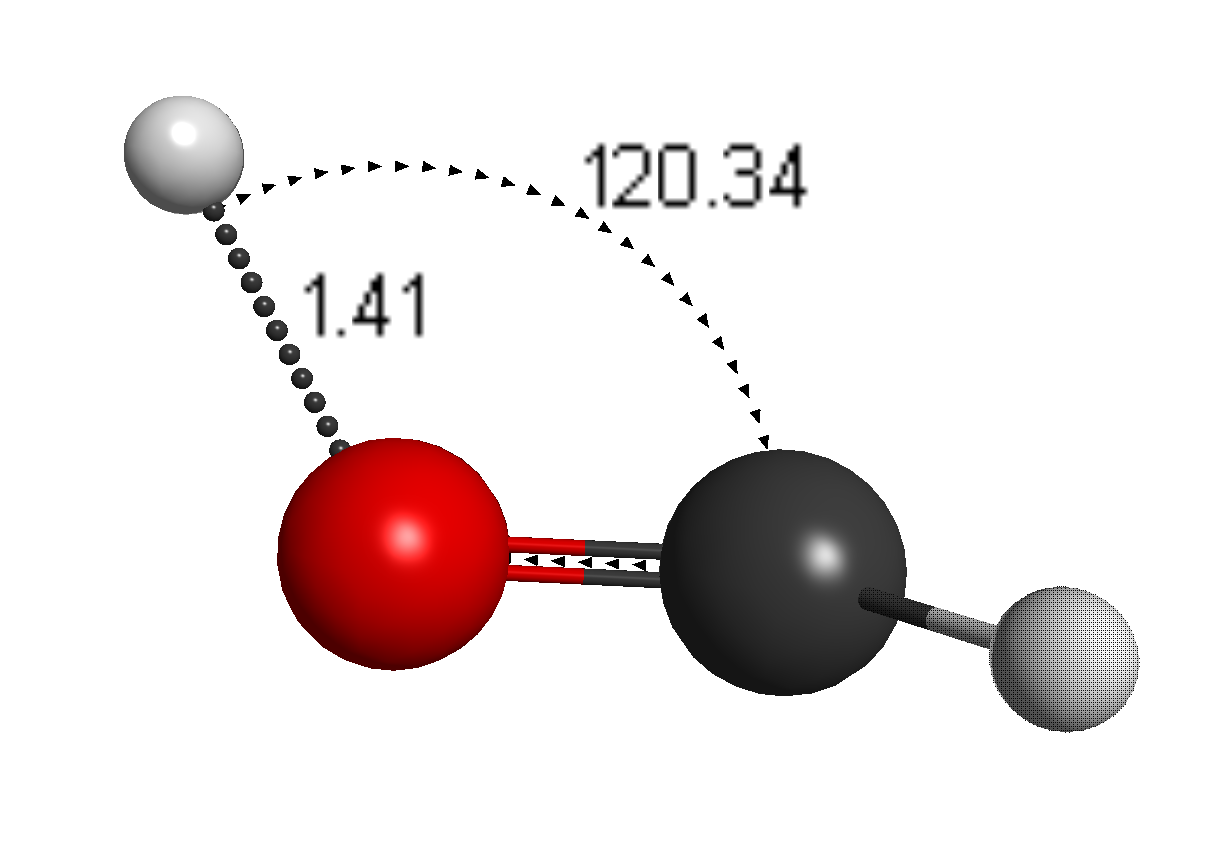} &
        $0.218$  & $2.20$ \\
        TS3 &
        \includegraphics[width=0.2\linewidth,valign = c]{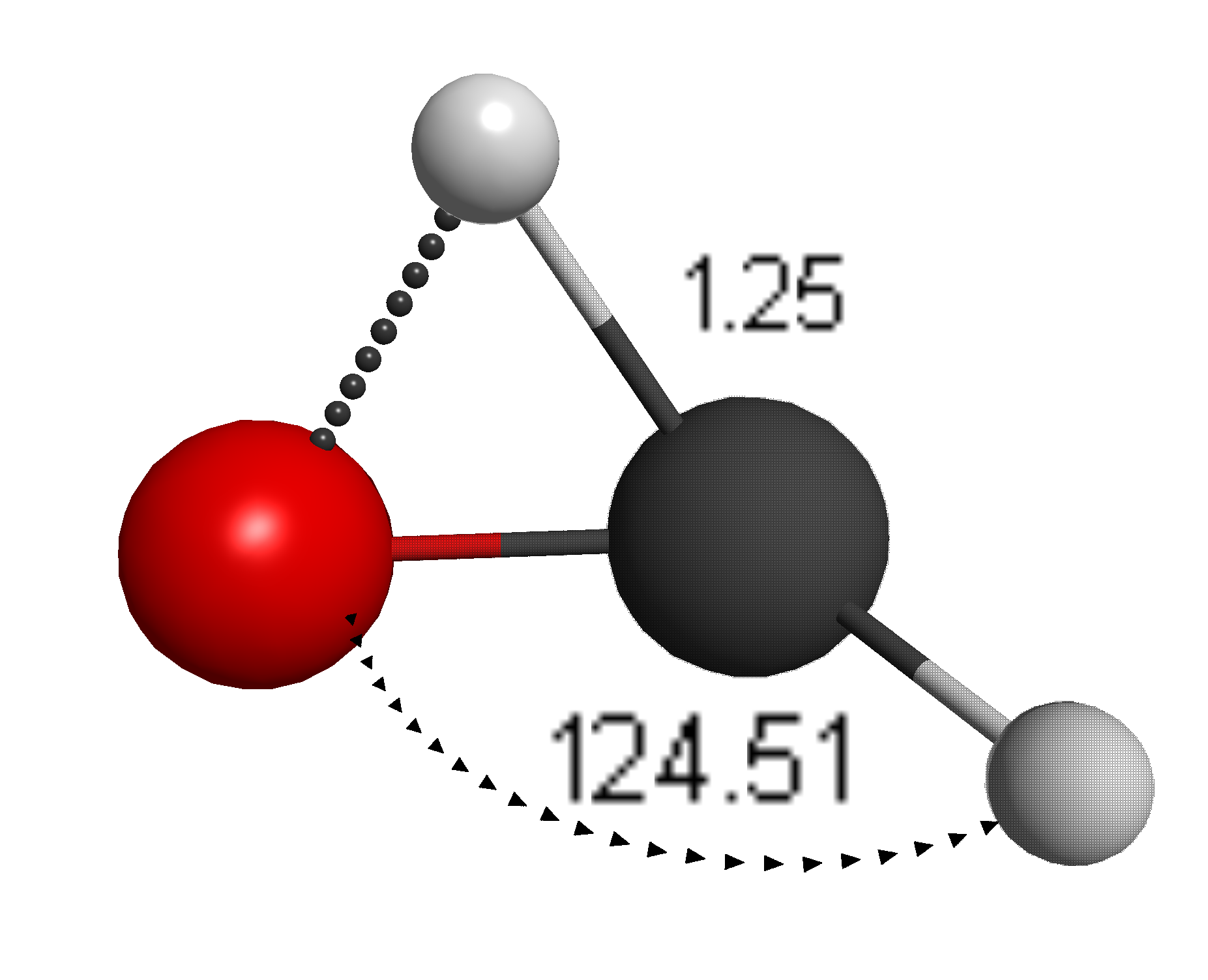} &
        \includegraphics[width=0.2\linewidth,valign = c]{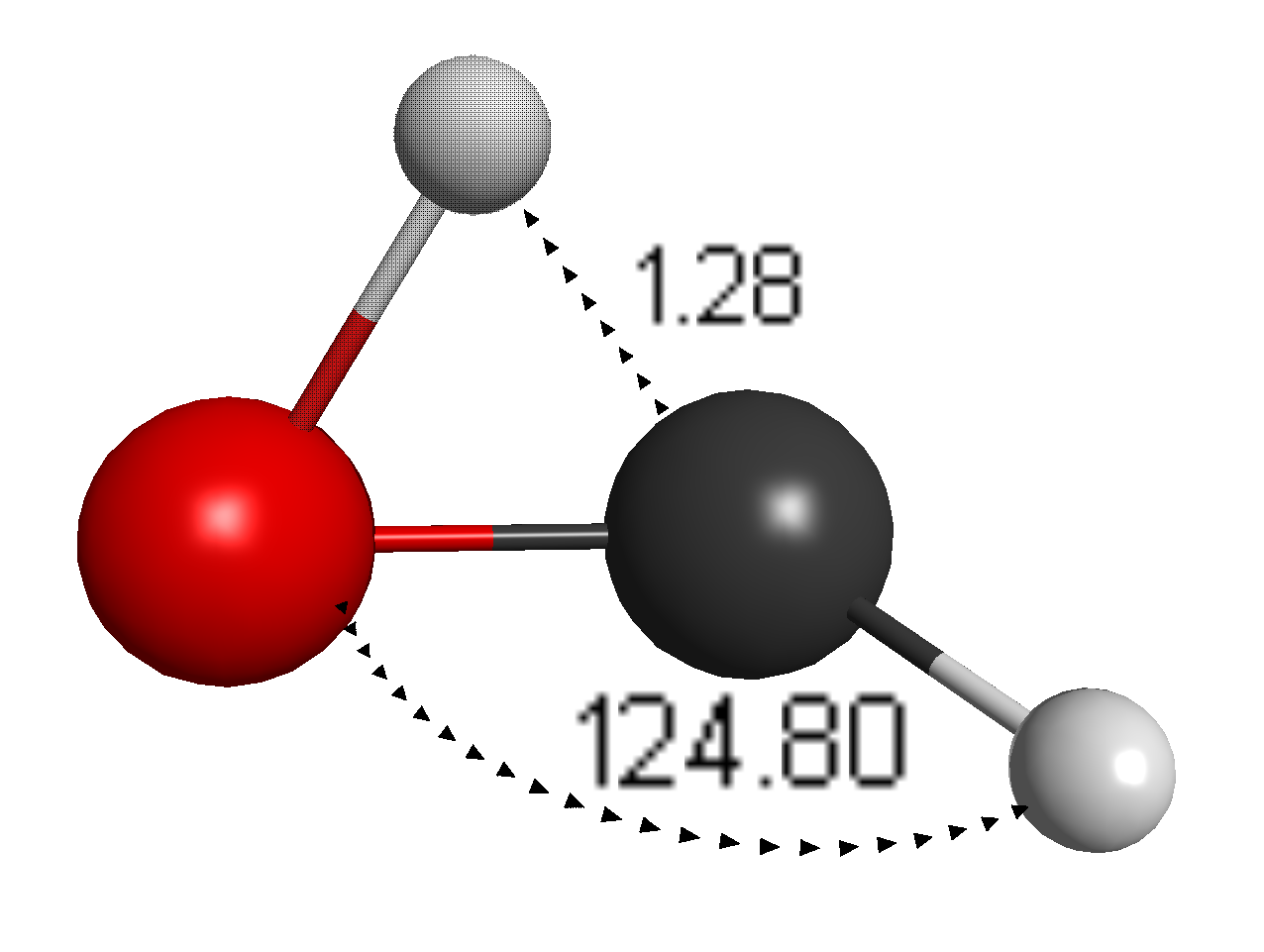} &
        $0.039$ & $0.87$ \\
        SOSP & 
        \includegraphics[width=0.2\linewidth,valign = c]{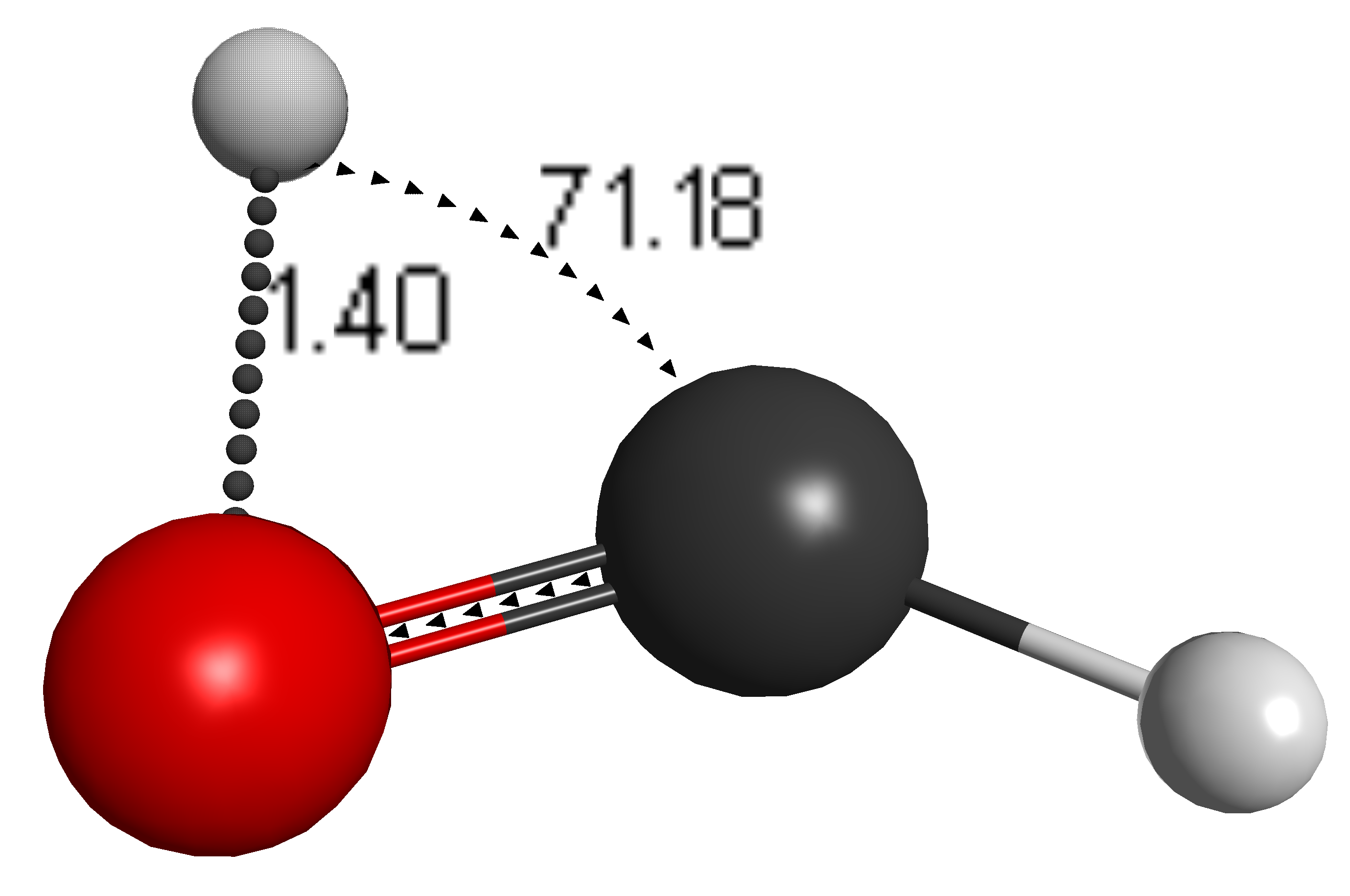} &
        \includegraphics[width=0.2\linewidth,valign = c]{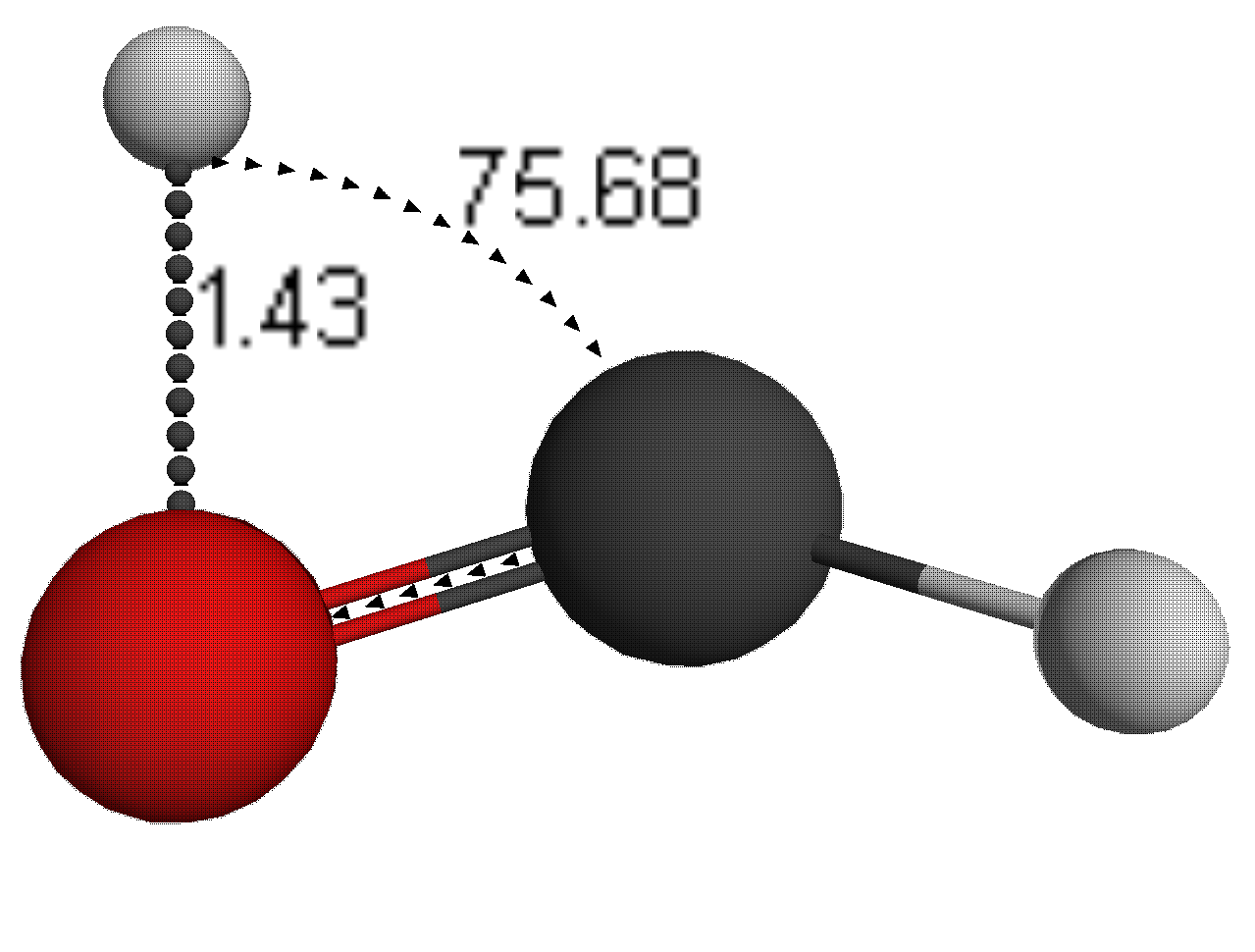} & $0.125$ & $0.070$\\
        \hline
        Mean & & &$0.150$ & $1.485$\\
        \bottomrule
    \end{tabular}
    \label{tab:H2CO_TSs}
\end{table*}

\section{Discussion and Outlook}

In this work, we introduced the nudged elastic membrane method as a framework for constructing and analyzing two-dimensional reduced potential-energy surfaces for chemical reactions. Unlike one-dimensional path methods, the NEM method is designed to capture the connectivity of the PES, including multiple reaction channels, transition-state regions, and higher-order features such as SOSPs. The method was demonstrated on a three-dimensional prototype model and on the triplet formaldehyde reaction network. In both cases, the resulting membranes recovered chemically meaningful features of the PES and provided useful approximate structures that can serve as starting points for more accurate local refinement and transition-state searches. Conceptually, the NEM method may be viewed as a practical procedure to construct an approximate minimum energy surface, i.e. a two-dimensional reduced manifold intended to retain the relevant reactive connectivity of the underlying PES while minimizing orthogonal force components. In this sense, the method is useful not only for identifying pathways and stationary structures, but also for defining a reduced surface on which reduced-dimensional dynamical approaches may be formulated.

The present implementation should be viewed as an  initial confirmation of feasibility. 
The quality of the resulting membrane depends on the initialization, grid resolution, and optimization protocol. 
Further work is needed to determine how strongly the results depend on the choice of tangent definition and minimization strategy~\cite{makri2019preconditioning,trygubenko2004doubly} and the use of non-cartesian coordinate systems such as non-redundant coordinates~\cite{zimmerman2013growing,schlegel2011geometry,henkelman2000climbing}. 
Even with these limitations, the present results show that the NEM framework is a promising route toward reduced-dimensional representations of the PES and toward efficient identification of chemically important structures on complex PESs.

\bibliography{references}

\newpage

\clearpage
\setcounter{equation}{0}
\setcounter{figure}{0}
\setcounter{table}{0}
\setcounter{section}{0}
\setcounter{page}{1}
\makeatletter
\renewcommand{\theequation}{S\arabic{equation}}
\renewcommand{\thefigure}{S\arabic{figure}}
\onecolumngrid
\section{Supplementary Information}
In this SI, we provide additional information about the two models studied.
\subsection{3D Model}
\begin{figure*}[h]
    \centering
    \includegraphics[width=0.65\linewidth]{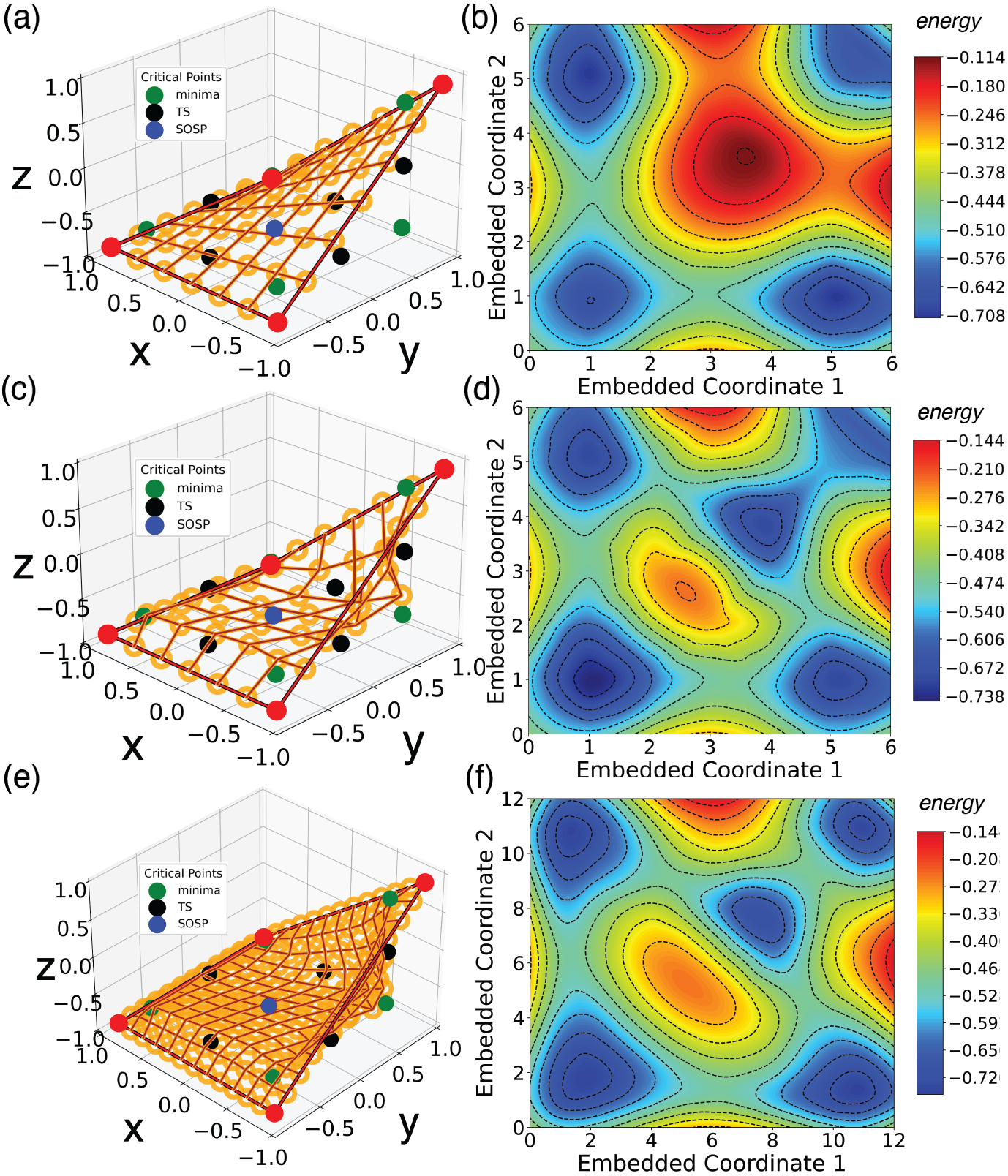}
   \caption{Visualization of NEM grid relaxation for the initial $7\times7$ (a-d) and $13\times13$ (e\&f) for the 3D model (Eq.~\ref{eq:3dmodel}). Initial grid (a) and energy surface (b) before optimization. Final (c) grid and (d) energy surface after optimization.
   Final (e) grid and (f) energy surface after optimization for the $13\times13$.
   Red points mark the fixed anchor beads, blue points represent SOSPs and black points indicate TSs.}
    \label{figS:3D_dbw_Grid}
\end{figure*}

Fig.~\ref{figS:3D_dbw_Grid} displays the chosen initial grid and energy surface of the 3D model discussed in the main text, and displays the convergence of the surface with respect to grid size.
\subsection{Triplet Formaldehyde}
\begin{figure*}[h]
    \centering
    \includegraphics[width=0.65\linewidth]{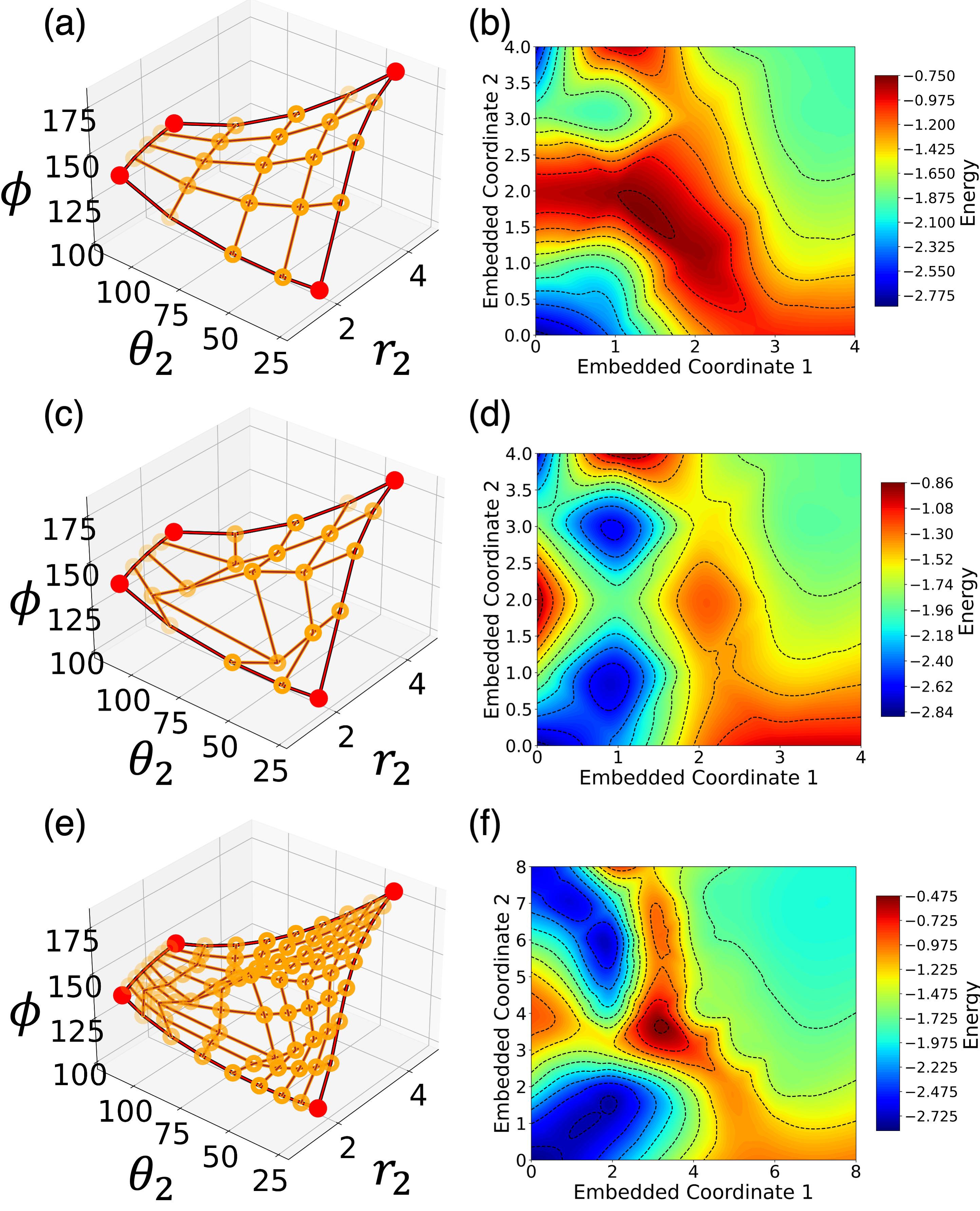}
   \caption{Visualization of NEM grid relaxation and upscaling of $5\times5$ (a-d) and $9\times9$ (e\&f) grids for the formaldehyde system. We use a different set of internal coordinates than shown in Fig.~\ref{fig:init_grid} of the main text to show the convergence more clearly. Initial grid (a) and energy surface (b) before optimization. 
   Final (c) grid and (d) energy surface after optimization.
   Final (e) grid and (f) energy surface after optimization for the $9\times9$ grid.}
    \label{figS:H2CO_upscale}
\end{figure*}

\begin{figure*}[h]
    \centering
    \includegraphics[width=0.65\linewidth]{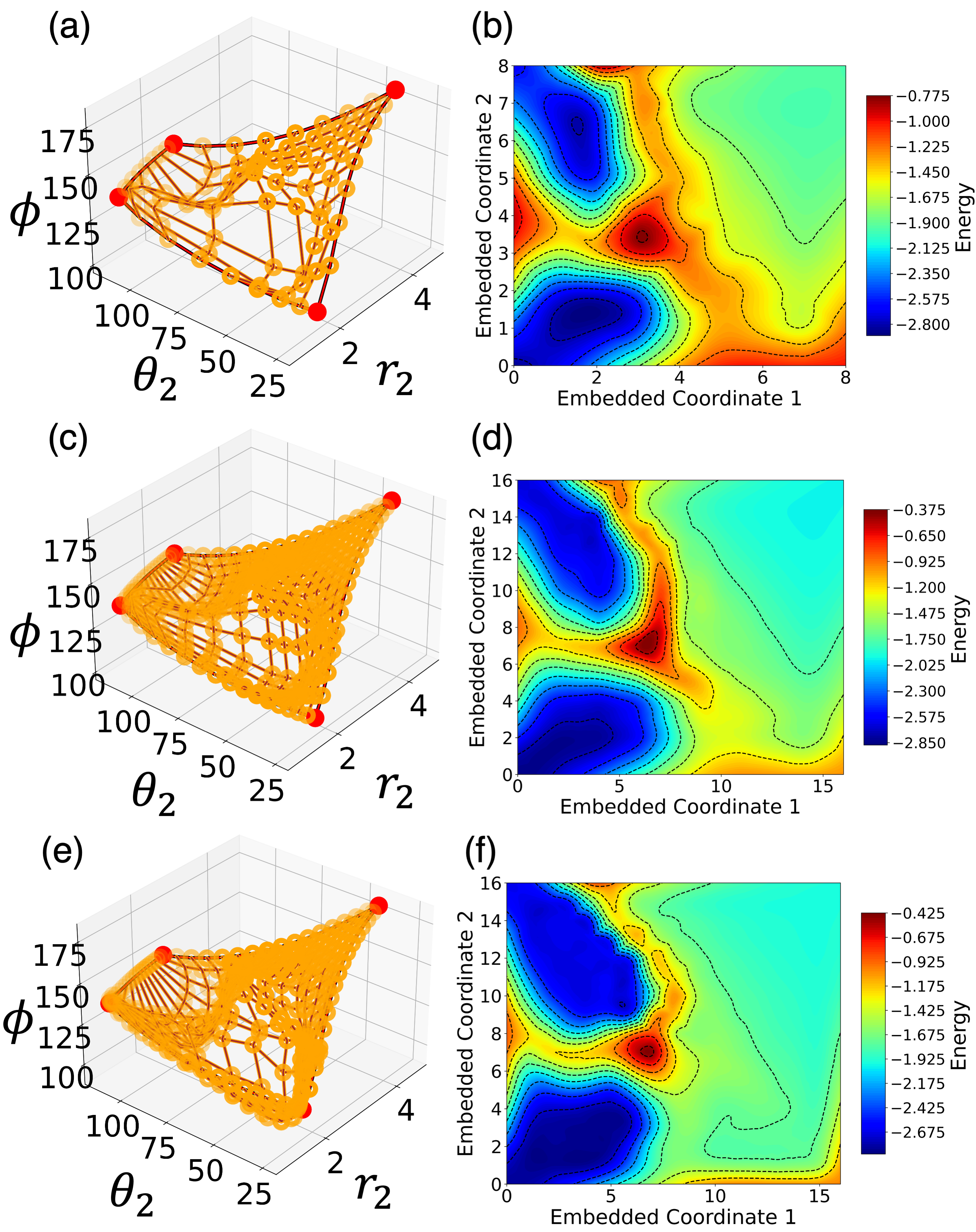}
   \caption{Visualization of NEM grid relaxation and upscaling of $9\times9$ (a\&b) and $17\times17$ (c-f) grids for the formaldehyde system. We use a different set of internal coordinates than shown in Fig.~\ref{fig:init_grid} of the main text to show the convergence more clearly. The optimized $9\times9$ grid and corresponding energy surface (b). 
   The initial $17\times17$ (c) grid and (d) energy surface before optimization.
   Final (e) grid and (f) energy surface after optimization for the $17\times17$ grid.}
    \label{figS:H2CO_upscale_2}
\end{figure*}

Fig.~\ref{figS:H2CO_upscale} visualizes the optimization and upscaling of the initial $5\times5$ grid along the internal coordinates ($r_2$,$\theta_2$, $\phi$). 
Here, $\phi$ is the dihedral angle angle $r_1$ and $\theta_2$ are the same internal coordinates shown in Fig.~\ref{fig:init_grid}. 
Fig.~\ref{figS:H2CO_upscale_2} visualizes the upscaling of the optimized $9\times9$ grid and the optimization of the $17\times17$ grid.

\end{document}